\documentclass[conference]{IEEEtran}  
\usepackage[dvips]{graphicx}
\usepackage{epsfig,amssymb,amsmath,color,spconf}
\newcommand{\beq}{\begin{equation}}
\newcommand{\eeq}{\end{equation}}
\newcommand{\beqn}{\begin{eqnarray}}

\newcommand{\eeqn}{\end{eqnarray}}
\def\bmath#1{\mbox{\boldmath$#1$}}
\DeclareMathOperator*{\argmax}{arg\,max}

\long\def\symbolfootnote[#1]#2{\begingroup%
\def\thefootnote{\fnsymbol{footnote}}\footnote[#1]{#2}\endgroup}

\title{Radio Astronomical Image Deconvolution Using Prolate Spheroidal Wave Functions}
\name{Sarod Yatawatta}
\address{
Kapteyn Astronomical Institute, University of Groningen, Groningen,\\
and ASTRON, Dwingeloo,\\
 The Netherlands.}

\begin{document}
\ninept \maketitle
%

\begin{abstract}
In order to produce high dynamic range images in radio interferometry, bright extended sources need to be removed with minimal error. However, this is not a trivial task because the Fourier plane is sampled only at a finite number of points. The ensuing deconvolution problem has been solved in many ways, mainly by algorithms based on CLEAN. However, such algorithms that use image pixels as basis functions have inherent limitations and by using an orthonormal basis that span the whole image, we can overcome them \cite{SBY10}. The construction of such an orthonormal basis involves fine tuning of many free parameters that define the basis functions. The optimal basis for a given problem (or a given extended source) is not guaranteed. In this paper, we discuss the use of generalized prolate spheroidal wave functions as a basis. Given the geometry (or the region of interest) of an extended source and the sampling points on the visibility plane, we can construct the optimal basis to model the source. Not only does this gives us the minimum number of basis functions required but also the artifacts outside the region of interest are minimized.
\end{abstract}

\section{Introduction}
Due to the increase in computing power, high dynamic range imaging  in radio interferometry (only limited by calibration errors) is achievable and is essential to produce novel scientific results. One of the main obstacles to reach a high dynamic range is the fact that only a finite number of samples of the Fourier (visibility) plane data is observed. Moreover, in earth rotation synthesis, these sampling points are not on a regular grid. If the observed field of view contains bright extended sources, complete removal (deconvolution) of such sources to reveal the faint background remains a challenge. The commonly used algorithm for such problems is CLEAN \cite{Hogbom}. However, as shown in \cite{SBY10} (and references therein), the usage of a set of image pixels (clean components) in CLEAN based algorithms has limitations. 

An alternative approach to using clean components is to use an orthonormal basis to model bright extended structure in the observation. Once such a model is obtained, it can be subtracted from the visibilities to reveal the residual (or the fainter background). The most obvious method of deriving such a basis is to use the observed data itself \cite{Levanda}. However, this has the drawback that the derivation of the basis can only be done once the complete observation is available (problematic for real time imaging etc.). On the other hand, we can adopt any arbitrary orthonormal basis to a given observation, completely independent of the data. For instance, in \cite{SBY10} Gauss-Hermite polynomials (shapelets) were used to produce high dynamic range images. The drawback in such an approach is the selection of free parameters (such as the number of basis functions, and the scale) cannot be determined in an optimal fashion. It requires experience of the user as well as a trial and error approach to fine tune such a basis for a given observation. Due to the noise floor, any extended source has finite support \cite{Slepian}, and the region of interest (ROI) or the support of a given extended source might not be optimal for a given arbitrary basis.

In order to tackle the problem of finding the optimal basis for a given extended source, independent of the observed data, we select prolate spheroidal wave functions (PSWF) \cite{Slepian61,Landau}. A similar problem has been solved in magnetic resonance imaging \cite{Lindquist06} and we extend that result to radio interferometry in this paper. Unlike data derived basis functions, PSWF basis can be precomputed and can be reused for observations at different epochs. Furthermore, unlike shapelets, we suffer less from artifacts outside the ROI with minimal number of basis functions used. In fact, PSWF are already being used in radio astronomical imaging to construct a regular grid of sampling points in the Fourier plane \cite{Brouw75}. We refer the reader to \cite{Simons} for similar applications of PSWF in geoscience.

Notation: We denote vectors in bold lowercase and matrices in bold uppercase. The matrix transpose, Hermitian, pseudoinverse are denoted by $(.)^T$, $(.)^H$ and $(.)^\dagger$ respectively. The identity matrix is given by ${\bf I}$.
\section{Mathematical Foundations}
We present the basics of interferometric imaging and the use of PSWF in this section. For a complete overview of radio interferometry, the reader is referred to \cite{Brouw75}.
\subsection{Interferometric imaging}
\begin{figure}[htb]
\begin{center}
 \input{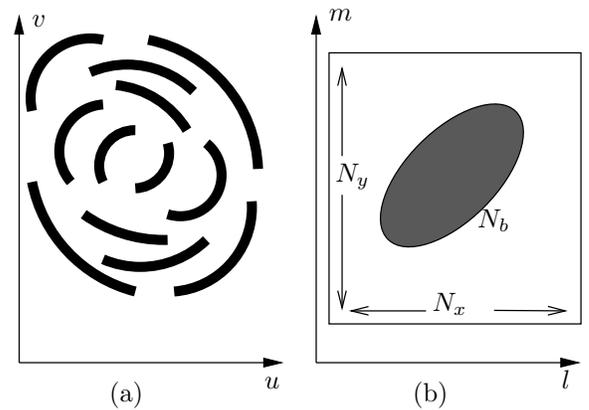}
\caption{(a) Sampling points (total $N_a$) in the Fourier (visibility) plane. (b) Image of $N_x$ by $N_y$ pixels, with the support (ROI) area shaded. The shaded area has $N_b$ pixels.\label{fig:roi}}
\end{center}
\end{figure}

We consider the visibility plane to be composed of $N_a$ sampling points as in Fig. \ref{fig:roi} (a). The image has $N$ ($=N_x \times N_y$) pixels. However, the ROI has only $N_b$ pixels, corresponding to the shaded area in Fig. \ref{fig:roi} (b). The ROI is determined by the support of the source structure and the noise floor \cite{Slepian}. 

Let $\widetilde{f}(u_p,v_p)$ be the sampled visibility at the $p$-th point in the visibility plane. Let us also denote the intensity of the $q$-th pixel on the image as $f(l_q,m_q)$. These two quantities are related by the van Cittert-Zernike theorem \cite{Brouw75} and can be approximated by the Fourier transform for images with small support as:
\beqn \label{fourier}
f(l_q,m_q)=\sum_{p=0}^{N_a-1} \widetilde{f}(u_p,v_p) e^{j 2 \pi (l_q u_p+m_q v_p)},\ q\in[0,N-1]\\\nonumber
\widetilde{f}(u_p,v_p)=\sum_{q=0}^{N-1} f(l_q,m_q) e^{-j 2 \pi (l_q u_p+m_q v_p)},\ p\in[0,N_a-1].
\eeqn
We can represent (\ref{fourier}) in vectorized form as
\beqn \label{vecfourier}
\widetilde{\bf f}&=&{\bf T} {\bf f},\ \ {\bf f}={\bf T}^{H} \widetilde{\bf f}\\\nonumber
{\rm where}\ \ {\bf f}&\buildrel \triangle \over =&[f(l_0,m_0),\ldots,f(l_{N-1},m_{N-1})]^T,\\\nonumber
 \widetilde{\bf f}&\buildrel \triangle \over =&[\widetilde{f}(u_0,v_0),\ldots,\widetilde{f}(u_{N_a-1},v_{N_a-1})]^T.
\eeqn
 The  matrix ${\bf T}$ (size $N_a\times N$) has on its $q$-th row and $p$-th column $e^{-j 2 \pi (l_p u_q+m_p v_q)}$. Note that given $\widetilde{\bf f}$, an estimate for the true image can be formed as $\widehat{\bf f}={\bf T}^{H}\widetilde{\bf f}$ but we recover the true image only if ${\bf T}^{H}{\bf T}={\bf I}$. In order to satisfy this condition we need a regular grid of sampling points in the visibility plane. This is not the case in reality for radio interferometry because the sampling points are determined by location of individual receivers and earth rotation.

\subsection{Prolate spheroidal basis}
Our objective is to find a basis function $p(l,m)$ (and its counterpart in the visibility plane $\widetilde{p}(u,v)$)  that maximizes the energy in the ROI. The criterion for selection can be written as
\beq \label{lambda}
\lambda = \frac{\sum_{(l,m)\in {\mathrm{ROI}}} |p(l,m)|^2}{\sum_{(l,m)} |p(l,m)|^2}.
\eeq
In other words, $\lambda$ in (\ref{lambda}) is the ratio between the energy concentrated in the ROI and the total energy in the image. The higher the value of $\lambda$, we have lower sidelobes and artifacts outside the ROI.

Let the vectorized versions of $p(l,m)$ and $\widetilde{p}(u,v)$, evaluated at the pixels and visibility points, be ${\bf p}$ and $\widetilde{\bf p}$, respectively. Then, we can rewrite (\ref{lambda}) as
\beq \label{lambdad}
\lambda = \frac{ \| {\bf I}_b^T {\bf p} \|^2}{\|{\bf p}\|^2}
\eeq

The selection of pixels that belong to the ROI from the full image vector ${\bf p}$ is done by premultiplying by ${\bf I}_b^T$. Thus, ${\bf I}_b^T$ is an $N_b\times N$ matrix, constructed by removing rows (corresponding to pixels outside the ROI) from an $N\times N$ identity matrix. 

We state the main result here, which is a direct extension of \cite{Lindquist06}. The steps required for the removal of an extended source from a given observation are:
\begin{enumerate}
\item Construct the kernel ${\bf K}$ (size $N_b$ by $N_b$) as
\beq \label{kernel}
{\bf K}={\bf I}_b^T{\bf T}^H ({\bf T} {\bf T}^H)^\dagger {\bf T}{\bf I}_b
\eeq
\item Find the eigendecomposition of ${\bf K}$ and select the eigenmodes with largest eigenvalues.  If the $i$-th eigenvalue, eigenvector pair of ${\bf K}$ is $(\lambda_i,{\bmath \eta}_i)$, the $i$-th basis vector is
\beq \label{kbasis}
{\bf p}_i=\frac{1}{\lambda_i}{\bf T}^H({\bf T} {\bf T}^H)^\dagger {\bf T}{\bf I}_b {\bmath \eta}_i
\eeq
\item Represent the image as the vector ${\bf b}$ (size $N$ by 1). Decompose image ${\bf b}$,  using $M$ basis vectors ${\bf P}\buildrel\triangle\over=[{\bf p}_0,\ldots,{\bf p}_{M-1}]$
\beq \label{imdec}
 {\bf P}{\bf m}={\bf b},\ \ \widehat{\bf m}={\bf P}^\dagger {\bf b}
\eeq
to find the mode vector $\widehat{\bf m}$ (size $M$ by 1).
\item Find equivalent basis $\widetilde{\bf P}={\bf T}^H {\bf P}$ in the Fourier plane and subtract the model from the observed data {\bf z},
\beq \label{residual}
{\bf r}={\bf z}-\widetilde{\bf P} \widehat{\bf m}
\eeq
to get the residual data vector {\bf r}.
\end{enumerate}

The complete proof of the derivation of PSWF is given in \cite{Lindquist06}. For completeness, we give a sketch of proof as follows. The numerator of (\ref{lambdad}) can be written as
\beqn \label{numerator}
\| {\bf I}_b^T {\bf p} \|^2&\buildrel (a) \over =& \| {\bf I}_b^T {\bf T}^H \widetilde{\bf p} \|^2=trace( \widetilde{\bf p}^H {\bf T} {\bf I}_b {\bf I}_b^T {\bf T}^H \widetilde{\bf p})\\\nonumber
&\buildrel (b) \over =&trace({\bmath \zeta}^H {\bf R}^{-H}  {\bf T} {\bf I}_b {\bf I}_b^T {\bf T}^H {\bf R}^{-1} {\bmath \zeta} )\\\nonumber
\eeqn
Here, (a) is obtained by the substitution ${\bf p}={\bf T}^H \widetilde{\bf p}$ and
 (b) is obtained by using ${\bmath \zeta}\buildrel \triangle \over ={\bf R} \widetilde{\bf p}$, where ${\bf R}\buildrel \triangle \over= ({\bf T}{\bf T}^H)^{1/2}$. 
It is easy to simplify the denominator of (\ref{lambdad})  by substituting ${\bf I}_b={\bf I}$ in (\ref{numerator}) as
\beqn \label{denominator}
\| {\bf p} \|^2=trace({\bmath \zeta}^H{\bf R}^{-H}{\bf T}{\bf T}^H{\bf R}^{-1}{\bmath \zeta})\buildrel (c) \over=trace({\bmath \zeta}^H{\bmath \zeta})
\eeqn
We used the fact that ${\bf R}^{-H}{\bf T}{\bf T}^H{\bf R}^{-1}={\bf I}$ to obtain (c) in (\ref{denominator}).

Using (\ref{numerator}) and (\ref{denominator}), we can rewrite (\ref{lambdad}) as 
\beq \label{lambdak}
{\bmath \zeta}=\argmax_{{\bmath \zeta},\ \ \|{\bmath \zeta\|=1}} trace({\bmath \zeta}^H{\bf R}^{-H}{\bf T}{\bf I}_b{\bf I}_b^T{\bf T}^H{\bf R}^{-1}{\bmath \zeta})
\eeq
The solution to (\ref{lambdak}) is the largest eigenvalue,eigenvector pair of the matrix $\widetilde{\bf K}={\bf R}^{-H}{\bf T}{\bf I}_b{\bf I}_b^T{\bf T}^H{\bf R}^{-1}$. The dimension of $\widetilde{\bf K}$ ($N_a$ by $N_a$) makes the computation of eigendecomposition prohibitively expensive. In order to reduce this, we apply the transform ${\bmath \eta}\buildrel\triangle\over ={\bf I}_b^T {\bf T}^H {\bf R}^{-1}{\bmath \zeta}$ to $\widetilde {\bf K}  {\bmath \zeta}=\lambda {\bmath \zeta}$ to get
\beq \label{zeta}
{\bf I}_b^H{\bf T}^H{\bf R}^{-2}{\bf T}{\bf I}_b{\bmath \eta}=\lambda {\bmath \eta}
\eeq
which gives us the kernel ${\bf K}$ in (\ref{kernel}) which is of dimension $N_b$ by $N_b$. 

\subsection{Computational cost reduction}
Under the assumption $N_a\gg N>N_b$, the cost of computing (\ref{kernel}) involves finding the pseudoinverse $({\bf T} {\bf T}^H)^\dagger$ (size $N_a$ by $N_a$) whose rank is at most $N$. However, the rank of ${\bf K}$ will be at most $N_b$.  Instead of using ${\bf T}$, we downsample the visibility points to construct a matrix $\widehat{\bf T}$ of size $N_d$ by $N$ ($N_d\ge N$). This downsampling can be combined with the imaging weights used. Therefore, (\ref{kernel}) and (\ref{kbasis}) are evaluated using $\widehat{\bf T}$ instead of ${\bf T}$. However, when we calculate the residual in (\ref{residual}), we use the full matrix ${\bf T}$. 

\section{Example}
\begin{figure}[htbp]
\begin{minipage}{0.98\linewidth}
\begin{minipage}{0.98\linewidth}
\centering
 \centerline{\epsfig{figure=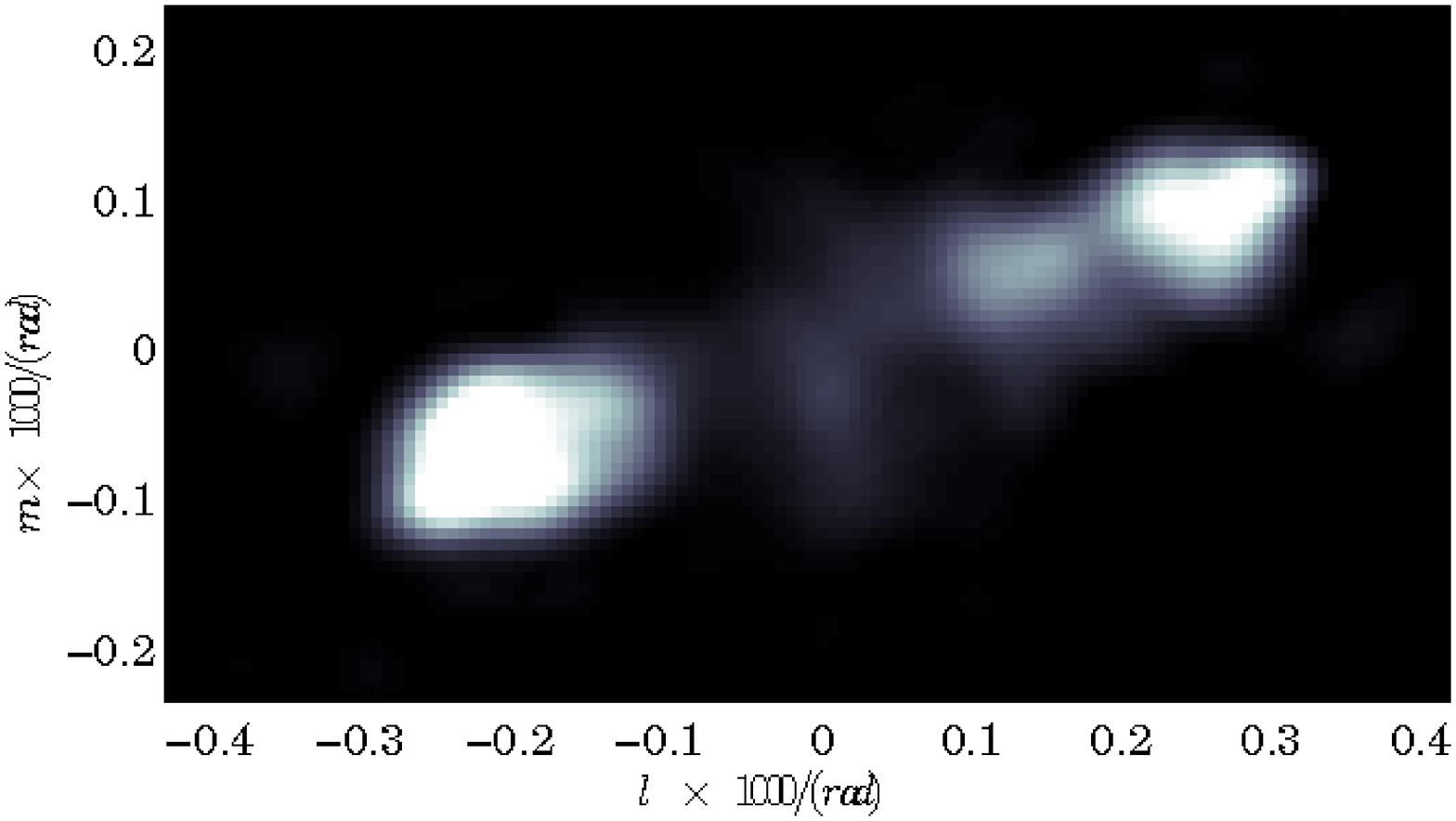,width=7.2cm}}
\vspace{0.1cm} \centerline{(a)}\smallskip
\end{minipage}\\
\begin{minipage}{0.98\linewidth}
\centering
 \centerline{\epsfig{figure=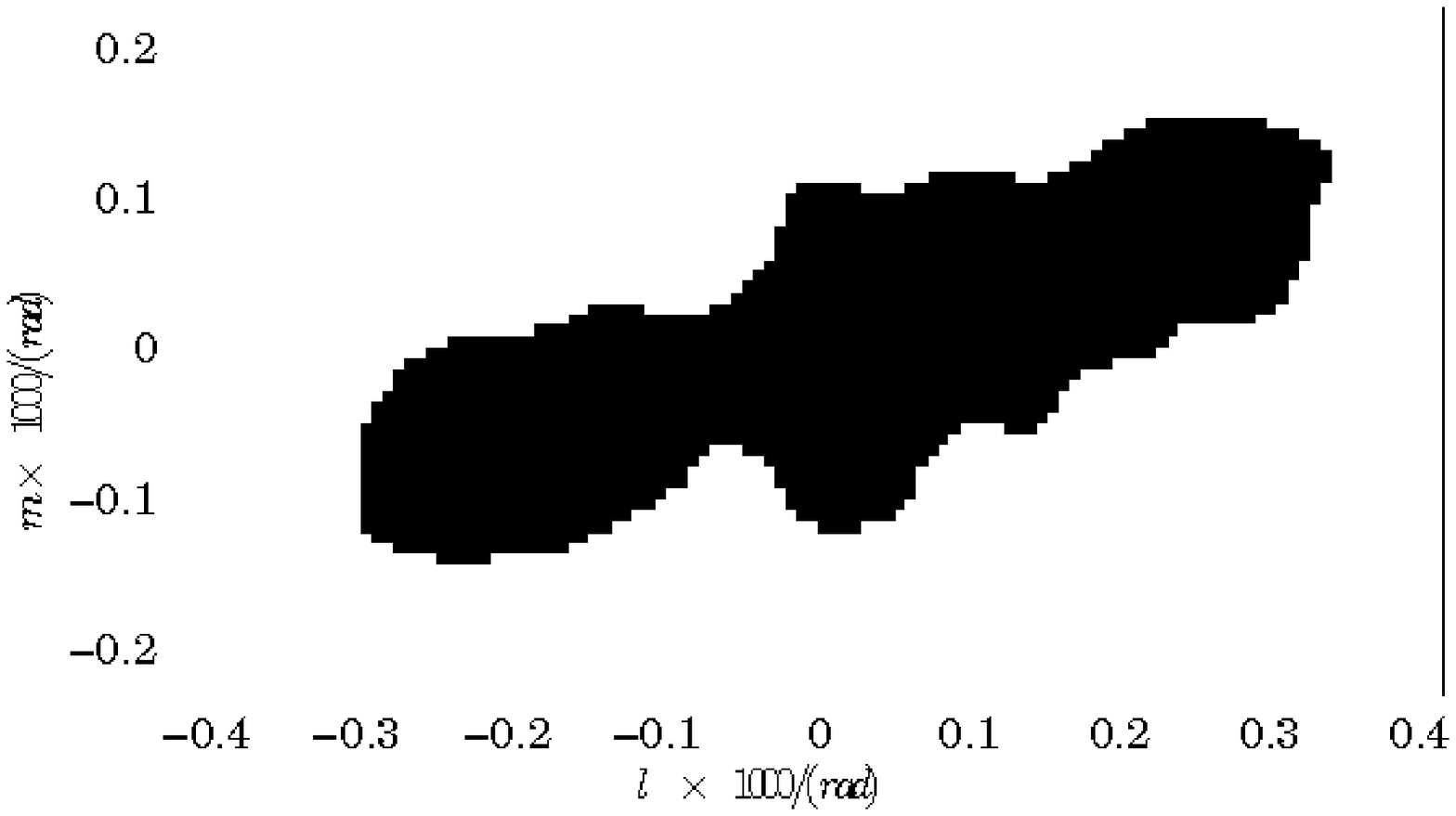,width=7.2cm}}
\vspace{0.1cm} \centerline{(b)}\smallskip
\end{minipage}
\end{minipage}
\caption{Cygnus A (a) Observed image at 213 MHz, deconvolved using CLEAN. The peak flux is about 20 Jy and to total flux is about 10 kJy.  (b) The shaded area correspond to the  ROI of 1826 pixels.\label{cyga}}
\end{figure}

We take a LOFAR ({http://www.lofar.org}) test observation of Cygnus A, at a frequency of 213 MHz as an example. The image of the source Cygnus A is given in Fig. \ref{cyga}(a). The peak flux (not normalized) is about 20 Jy and the total is about 10 kJy. The objective is to subtract this source from the observed data to see the faint background sources. The ROI of this source, which is above the noise floor is given in Fig. \ref{cyga} (b). The image has $N=7552$ pixels of dimension $118$ by $64$ and the ROI has $N_b=1826$ pixels.

\begin{figure}[htbp]
\begin{minipage}{0.99\linewidth}
\begin{minipage}{0.50\linewidth}
\centering
 \centerline{\epsfig{figure=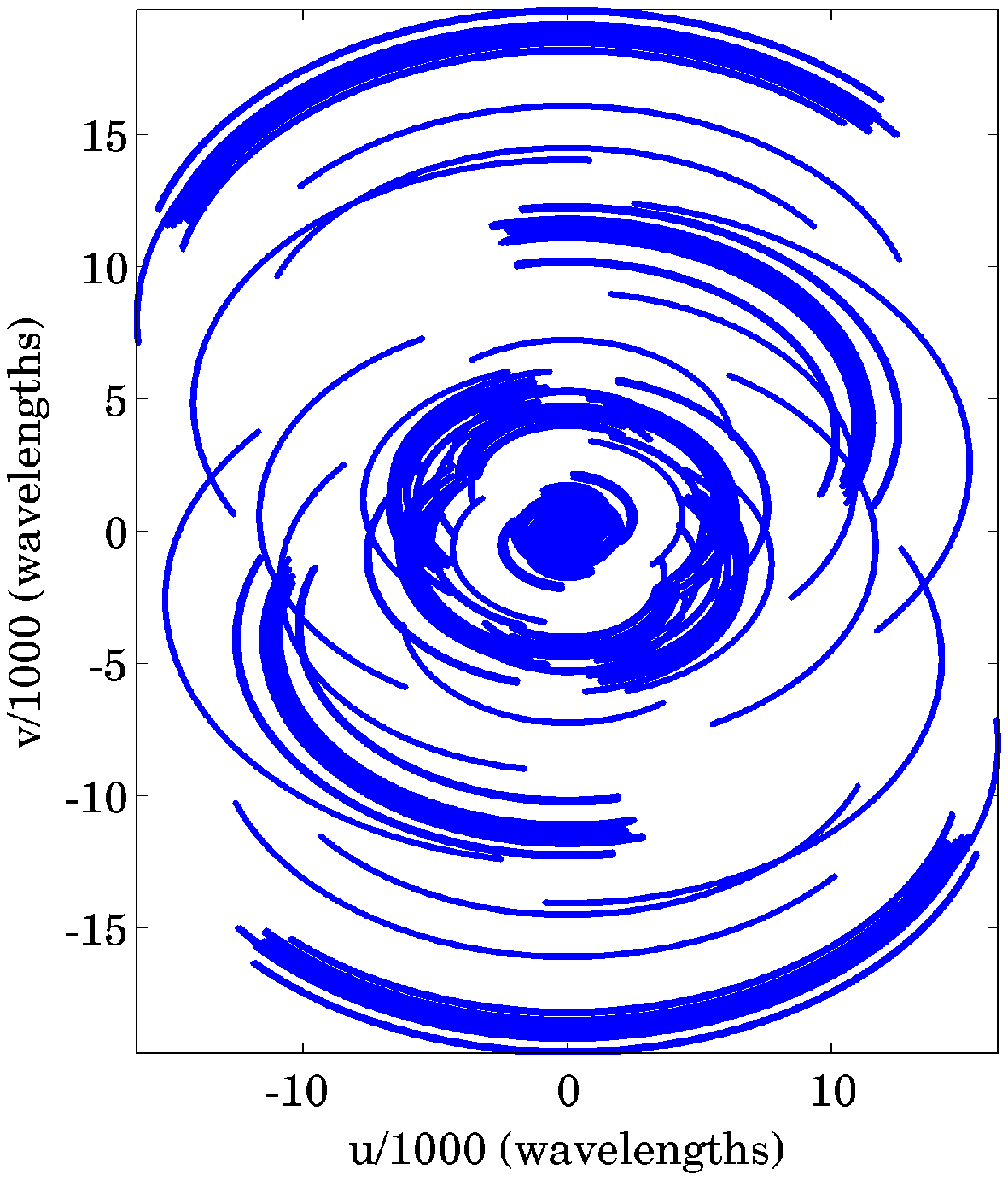,width=4.0cm}}
\vspace{0.1cm} \centerline{(a)}\smallskip
\end{minipage}
\begin{minipage}{0.50\linewidth}
\centering
 \centerline{\epsfig{figure=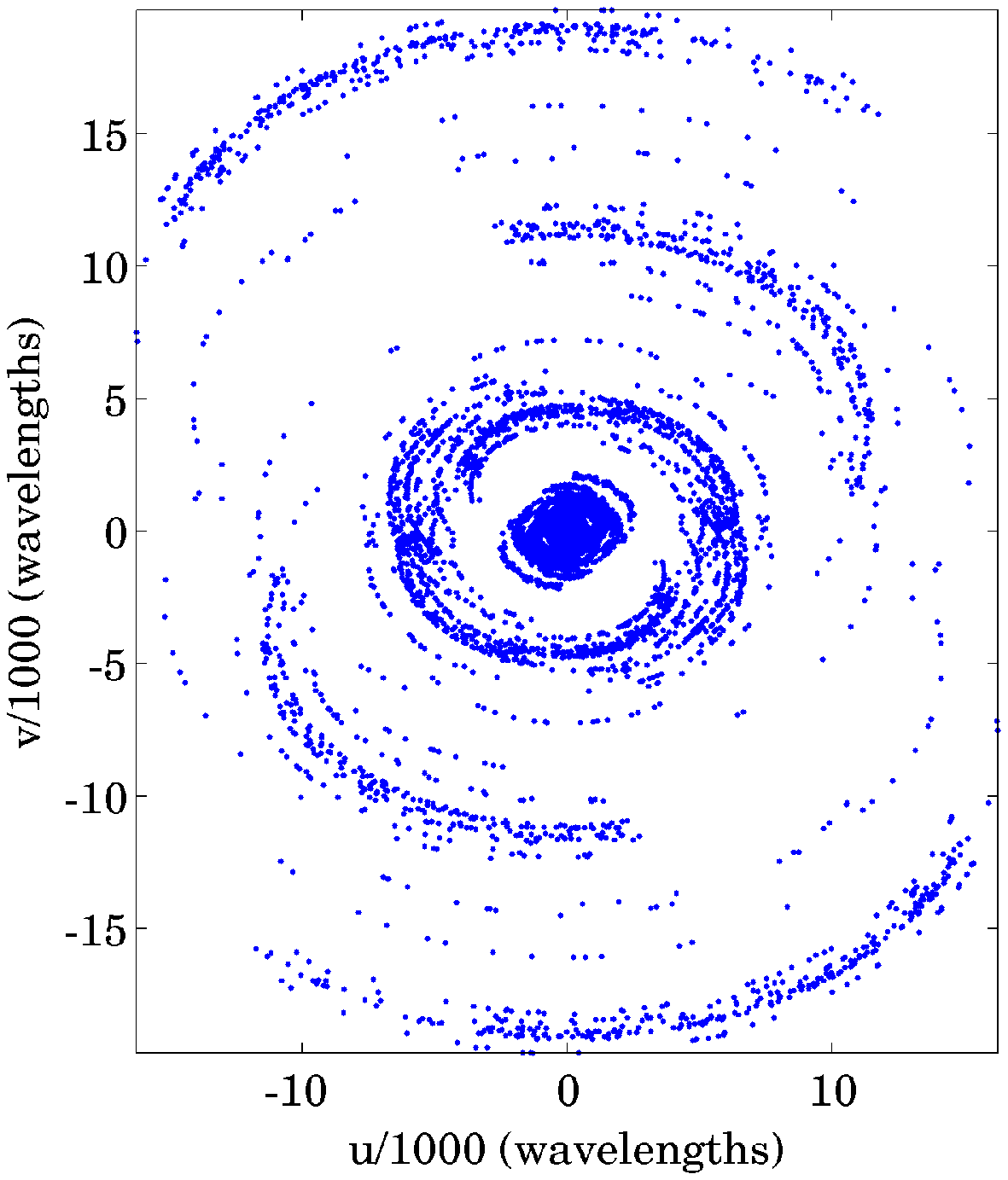,width=4.0cm}}
\vspace{0.1cm} \centerline{(b)}\smallskip
\end{minipage}
\end{minipage}
\caption{The sampling points in the Fourier plane (a) Original coverage, for an 8 hour observation with about 15 million sampling points. (b) Reduced coverage by downsampling by 1500 with 8931 sampling points. The downsampling is done with uniform probability but it is also possible to combine this with imaging weights, for instance by selecting fewer short baselines and more long baselines.\label{uvcov}}
\end{figure}

The observation lasted for about 8 hours and the original sampling points are shown in Fig. \ref{uvcov} (a). The number of sampling points is $N_a\approx 15\times 10^6$.  We downsample the sampling coverage (select only a subset with uniform probability) to get $N_d=8931$ which is just above $N$. A better approach for downsampling would be to combine this with imaging weights (i.e., selecting fewer short baselines and more long baselines).

\begin{figure}[htbp]
\begin{minipage}{0.98\linewidth}
\centering
 \centerline{\epsfig{figure=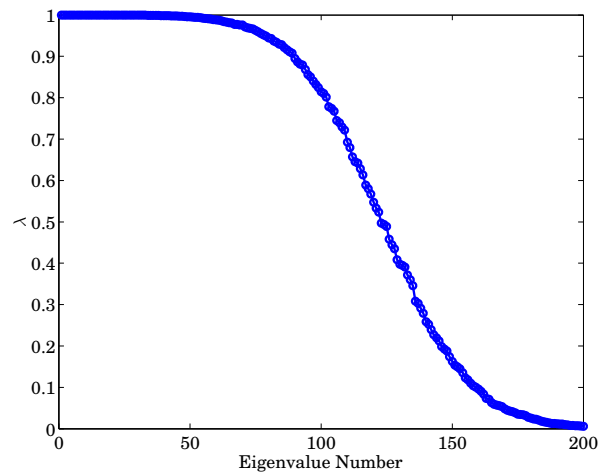,width=9.0cm}}
\end{minipage}
\caption{Eigenspectrum of the kernel ${\bf K}$ with the first 200 largest eigenvalues plotted.\label{eigenspectrum}}
\end{figure}

The dominant eigenvalues of the kernel ${\bf K}$ (\ref{kernel})  (dimension 1826 by 1826) is shown in Fig. \ref{eigenspectrum}. We see that approximately the first 100 eigenvalues are close to 1 while any eigenvalue beyond the 200-th eigenmode is almost zero.

\begin{figure}[htbp]
\begin{minipage}{0.98\linewidth}
\centering
 \centerline{\epsfig{figure=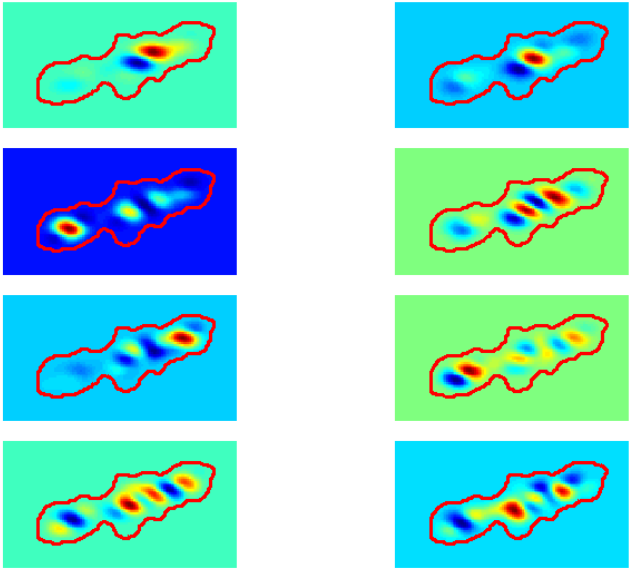,width=9.0cm}}
\end{minipage}
\caption{Some PSWF basis vectors corresponding to the largest eigenvalues. The ROI is indicated by the dark curve.\label{somebasis}}
\end{figure}

In Fig. \ref{somebasis}, we have shown the PSWF basis vectors ${\bf p}$ corresponding to the first few largest eigenvalues. Note that the support of the basis vectors are entirely within the ROI, thus creating minimal artifacts outside the ROI. We select  the first 100 basis functions $M=100$ to construct the matrix ${\bf P}$ in (\ref{imdec}). Using this, we decompose the image in Fig. \ref{cyga} (a) to get the mode vector ${\bf m}$.

\begin{figure}[htbp]
\begin{minipage}{0.98\linewidth}
\begin{minipage}{0.98\linewidth}
\centering
 \centerline{\epsfig{figure=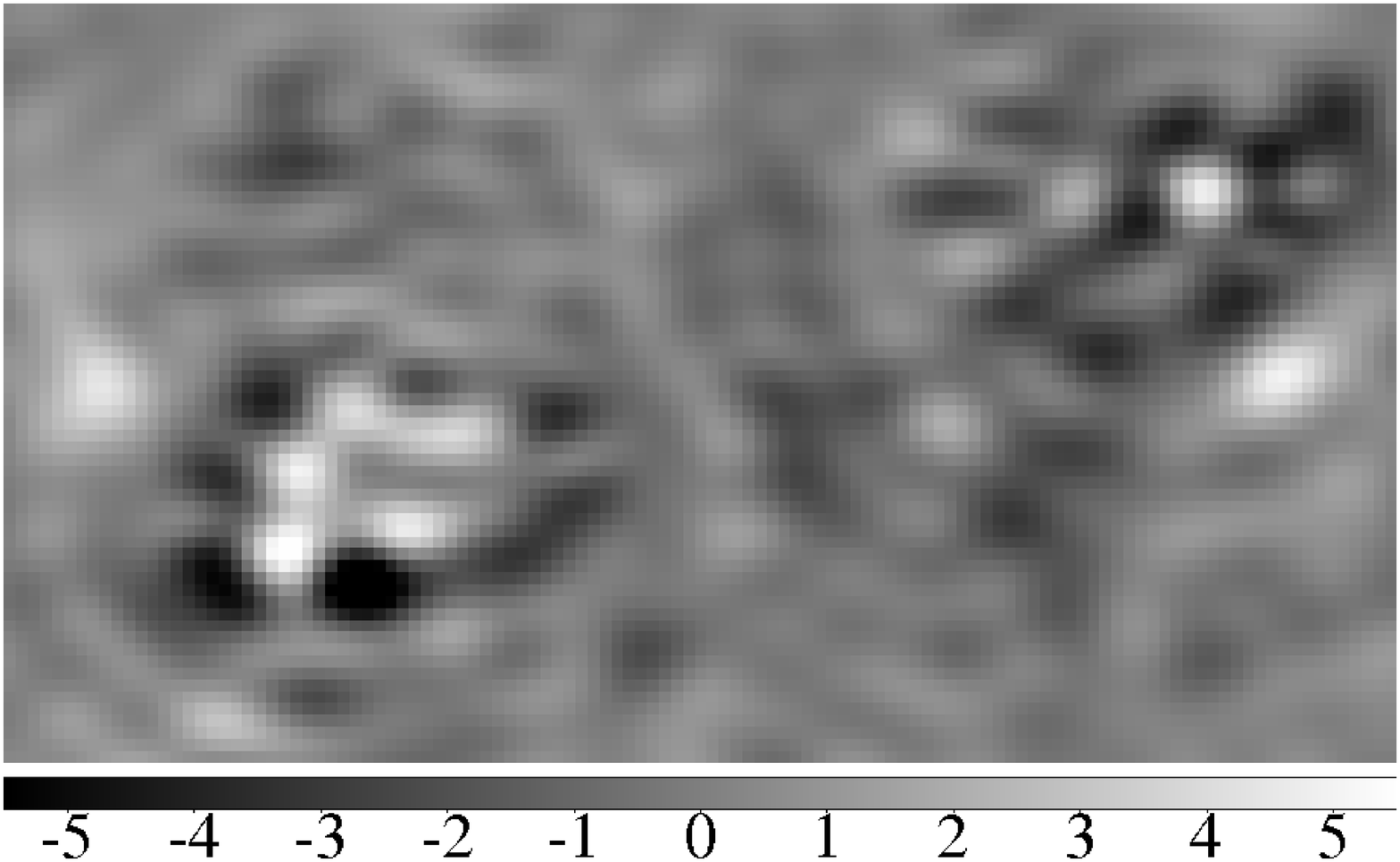,width=5.7cm}}
\vspace{0.1cm} \centerline{(a)}\smallskip
\end{minipage}\\
\begin{minipage}{0.98\linewidth}
\centering
 \centerline{\epsfig{figure=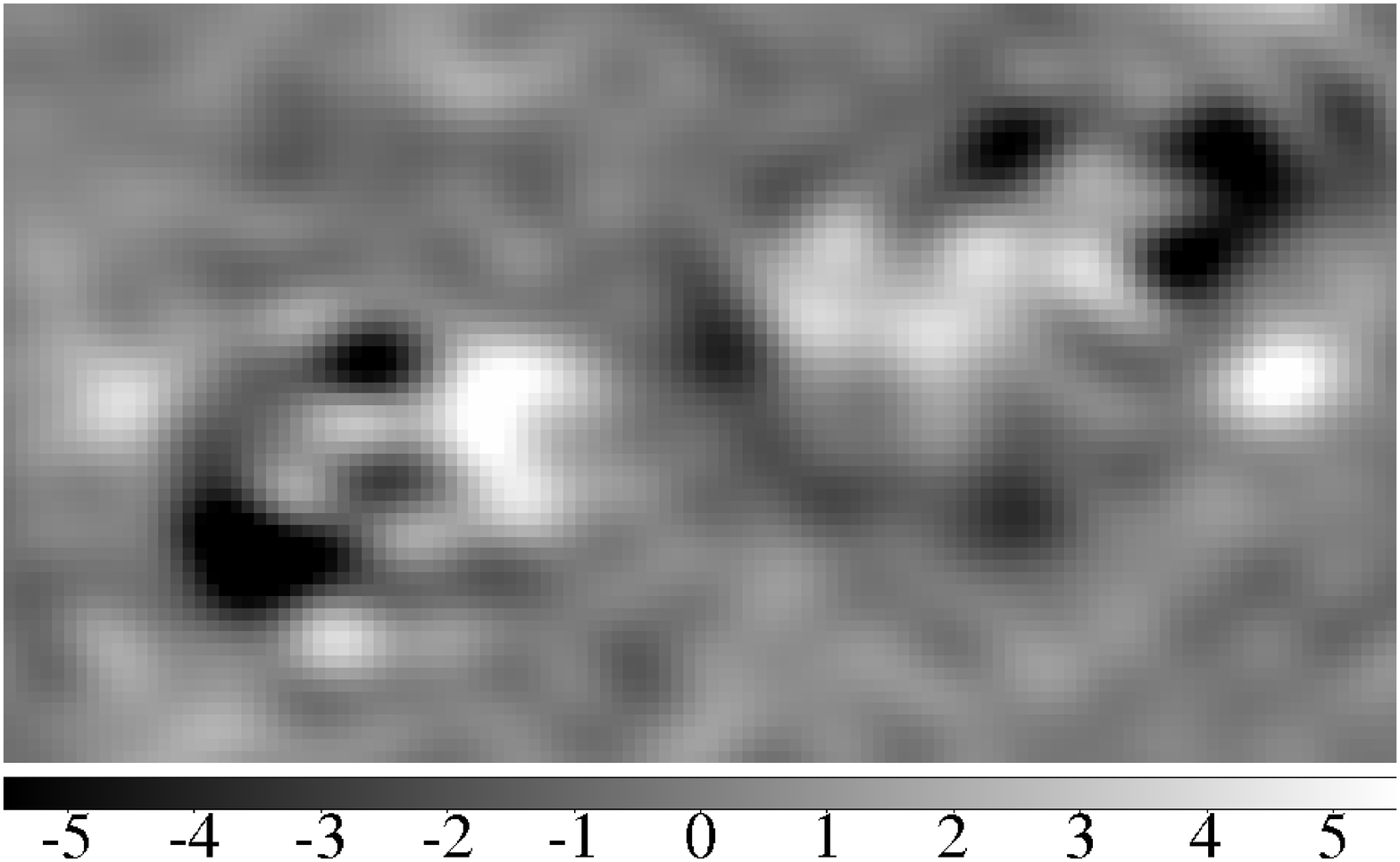,width=5.7cm}}
\vspace{0.1cm} \centerline{(b)}\smallskip
\end{minipage}
\end{minipage}
\caption{Residuals after subtracting Cygnus A using (a) prolate basis (b) shapelet basis. Note that in (b) there is more unsubtracted flux towards the center of the source while in (a) the residuals are compact (point like). Both images have peak values of 5.5 Jy.\label{residuals}}
\end{figure}

Using the equivalent basis in the Fourier plane, to calculate the residual as in(\ref{residual}). Once the residual is obtained we make the residual images as shown in Fig. \ref{residuals}. For comparison we have also shown the residual image obtained using a shapelet basis function based deconvolution. Both methods give a residual noise level of about 11 mJy far away from Cygnus A. In contrast, with traditional CLEAN based deconvolution, we get a residual noise of about 13 mJy. The shapelet based model used about 300 basis functions of many scales. On the other hand, the PSWF model used only about 100 modes, which is much less.

\section{Conclusions}
We have presented the use of prolate spheroidal wave functions in radio interferometric image deconvolution and have demonstrated its feasibility by application to a real observation. We get results comparable with existing techniques, but with fewer basis functions and with less artifacts outside the ROI. Future work will focus on widefield imaging and reducing the computational cost.
\section{Acknowledgments}
We thank Wim Brouw and Ger de Bruyn for valuable comments and advise.
\bibliographystyle{IEEEtran}
\bibliography{shapeletref}

\begin{thebibliography}{1}
\providecommand{\url}[1]{#1}
\csname url@samestyle\endcsname
\providecommand{\newblock}{\relax}
\providecommand{\bibinfo}[2]{#2}
\providecommand{\BIBentrySTDinterwordspacing}{\spaceskip=0pt\relax}
\providecommand{\BIBentryALTinterwordstretchfactor}{4}
\providecommand{\BIBentryALTinterwordspacing}{\spaceskip=\fontdimen2\font plus
\BIBentryALTinterwordstretchfactor\fontdimen3\font minus
  \fontdimen4\font\relax}
\providecommand{\BIBforeignlanguage}[2]{{%
\expandafter\ifx\csname l@#1\endcsname\relax
\typeout{** WARNING: IEEEtran.bst: No hyphenation pattern has been}%
\typeout{** loaded for the language `#1'. Using the pattern for}%
\typeout{** the default language instead.}%
\else
\language=\csname l@#1\endcsname
\fi
#2}}
\providecommand{\BIBdecl}{\relax}
\BIBdecl

\bibitem{SBY10}
S.~Yatawatta, ``{Fundamental limitations of pixel based image deconvolution in
  radio astronomy},'' \emph{in proc. IEEE Sensor Array and Multichannel Signal
  Processing Workshop (SAM), Israel}, pp. 69--72, 2010.

\bibitem{Hogbom}
J.~Hogbom, ``{Aperture synthesis with a non regular distribution of
  interferometer baselines},'' \emph{A\&A Suppl.}, vol.~15, pp. 417--426, 1974.

\bibitem{Levanda}
R.~Levanda and A.~Leshem, ``{Adaptive selective sidelobe canceller beamformer
  with applications in radio astronomy},'' \emph{in proc. IEEE 26-th Convention
  of Electrical and Electronics Engineers (IEEEI), Israel}, 2010.

\bibitem{Slepian}
D.~Slepian, ``{On bandwidth},'' \emph{Proc. of the IEEE}, vol.~64, no.~3, pp.
  292--300, 1976.

\bibitem{Slepian61}
D.~Slepian and H.~O. Pollak, ``{Prolate spheroidal wave functions, Fourier
  analysis, and uncertainty-I},'' \emph{Bell Syst. Tech. J.}, vol.~40, no.~2,
  pp. 43--61, 1961.

\bibitem{Landau}
H.~J. Landau and H.~O. Pollak, ``{Prolate spheroidal wave functions, Fourier
  analysis, and uncertainty-II},'' \emph{Bell Syst. Tech. J.}, vol.~40, no.~2,
  pp. 65--84, 1961.

\bibitem{Lindquist06}
M.~A. Lindquist, C.~Zhang, G.~Glover, L.~Shepp, and Q.~X. Yang, ``{A
  generalization of the two dimensional prolate spheroidal wave function method
  for nonrectiliner MRI data acquisition methods},'' \emph{IEEE Trans. on Image
  Proc.}, vol.~15, no.~9, pp. 2792--2804, 2006.

\bibitem{Brouw75}
W.~N. Brouw, ``{Aperture synthesis},'' \emph{in Methods in Computational
  Physics}, vol.~14, pp. 131--175, 1975.

\bibitem{Simons}
F.~J. Simons and D.~V. Wang, ``{Spatiospectral conventration in the Cartesian
  plane},'' \emph{Appl. Comput. Harmon. Anal. under review}, 2010.

\end{thebibliography}

\end{document}